# Approximation of a Fractional Order System by an Integer Order Model Using Particle Swarm Optimization Technique

Deepyaman Maiti and Amit Konar
Department of Electronics and Telecommunication Engineering, Jadavpur University, Kolkata - 700 032
E-mail: deepyamanmaiti@gmail.com, konaramit@yahoo.co.in

**Abstract**

*System identification is a necessity in control theory. Classical control theory usually considers processes with integer order transfer functions. Real processes are usually of fractional order as opposed to the ideal integral order models. A simple and elegant scheme is presented for approximation of such a real world fractional order process by an ideal integral order model. A population of integral order process models is generated and updated by PSO technique, the fitness function being the sum of squared deviations from the set of observations obtained from the actual fractional order process. Results show that the proposed scheme offers a high degree of accuracy.*

## 1. INTRODUCTION

Proper estimation of the parameters of a real process, fractional or otherwise, is a challenge to be encountered in the context of system identification [1], [2]. Accurate knowledge of the transfer function of a system is often the first step in designing controllers. Computation of transfer characteristics of the fractional order dynamic systems has been the subject of several publications, e.g. by numerical methods [3], as well as by analytical methods [4]. Many classical statistical and geometric methods such as least square and regression models are widely used for real-time system identification.

The problem of system identification becomes more difficult for a fractional order system compared to an integral order one. The real world objects or processes that we want to estimate are generally of fractional order [5] (for example, the voltage-current relation of a semi-infinite lossy RC line or diffusion of heat into a semi-infinite solid). The usual practice when dealing with such a fractional order process is to use an integer order approximation. In general, this approximation can cause significant differences between a real system and its mathematical model.

However, the concept of approximation of a real fractional order process by an integral order one is not without its merits, provided the approximation is sufficiently accurate. Classical control theory deals with integral order processes. A good approximation would enable us to analyze and control fractional order processes with the conventional theory.

In this paper, we propose a general method for the estimation of parameters of a fractional order system using PSO technique. PSO, a stochastic optimization strategy from the family of evolutionary computation, is a biologically-inspired technique originally proposed by Kennedy and Eberhart. PSO offers optimal or sub-optimal solution to multi-dimensional rough objective functions. We use this technique to find the integral order process model whose outputs match the set of observations from the actual fractional order system most closely. This method enables us to achieve a very good approximation.

It is necessary to understand the theory of fractional calculus in order to realize the significance of fractional order integration and derivation.

## 2. THEORY OF FRACTIONAL CALCULUS

The fractional calculus is a generalization of integration and derivation to non-integer order operators. At first, we generalize the differential and integral operators into one fundamental operator $_aD_t^\alpha$ where:

$$_aD_t^\alpha = \begin{cases} \dfrac{d^\alpha}{dt^\alpha}, & \Re(\alpha) > 0 \\ 1, & \Re(\alpha) = 0 \\ \displaystyle\int_a^t (d\tau)^{-\alpha}, & \Re(\alpha) < 0 \end{cases} \quad (1)$$

The two definitions used for fractional differintegral are the Riemann-Liouville definition [6] and the Grunwald-Letnikov definition.

The Riemann-Liouville definition is given as

$$_aD_t^\alpha f(t) = \frac{1}{\Gamma(n-\alpha)} \frac{d^n}{dt^n} \int_a^t \frac{f(\tau)}{(t-\tau)^{\alpha-n+1}} d\tau \quad (2)$$

for $(n-1 < \alpha < n)$ and $\Gamma(x)$ is Euler's gamma function.

The Grunwald-Letnikov definition is

$$_aD_t^\alpha f(t) = \lim_{h \to 0} \frac{1}{h^\alpha} \sum_{j=0}^{\left[\frac{t-a}{h}\right]} (-1)^j \binom{\alpha}{j} f(t-jh) \quad (3)$$

where $[x]$ means the integer part of x.
Derived from the Grunwald-Letnikov definition, the numerical calculation formula of fractional derivative can be achieved as:

$$_{t-L}D_t^\alpha x(t) \approx h^{-\alpha} \sum_{j=0}^{[L/T]} b_j x(t-jh) \qquad (4)$$

where L is the length of memory. T, the sampling time always replaces the time increment h during approximation. The weighting coefficients $b_j$ can be calculated recursively by:

$$b_0 = 1, b_j = \left(1 - \frac{1+\alpha}{j}\right) b_{j-1}, (j \geq 1). \qquad (5)$$

### 3. PARTICLE SWARM OPTIMIZATION TECHNIQUE

The PSO algorithm [7] - [10] attempts to mimic the natural process of group communication of individual knowledge, which occurs when a social swarm elements flock, migrate, forage, etc. in order to achieve some optimum property such as configuration or location.

The 'swarm' is initialized with a population of random solutions. Each particle in the swarm is a different possible set of the unknown parameters to be optimized. Representing a point in the solution space, each particle adjusts its flying toward a potential area according to its own flying experience and shares social information among particles. The goal is to efficiently search the solution space by swarming the particles toward the best fitting solution encountered in previous iterations with the intent of encountering better solutions through the course of the process and eventually converging on a single minimum error solution.

Let the swarm consist of N particles moving around in a D-dimensional search space. Each particle is initialized with a random position and a random velocity. Each particle modifies its flying based on its own and companions' experience at every iteration. The $i^{th}$ particle is denoted by $X_i$, where $X_i = (x_{i1}, x_{i2}, \ldots, x_{iD})$. Its best previous solution (pbest) is represented as $P_i = (p_{i1}, p_{i2}, \ldots, p_{iD})$. Current velocity (position changing rate) is described by $V_i$, where $V_i = (v_{i1}, v_{i2}, \ldots, v_{iD})$. Finally, the best solution achieved so far by the whole swarm (gbest) is represented as $P_g = (p_{g1}, p_{g2}, \ldots, p_{gD})$.

At each time step, each particle moves towards pbest and gbest locations. The fitness function evaluates the performance of particles to determine whether the best fitting solution is achieved. The particles are manipulated according to the following equations:

$$v_{id}(t+1) = \omega v_{id}(t) + c_1.\varphi_1.(p_{id}(t) - x_{id}(t)) + c_2.\varphi_2.(p_{gd}(t) - x_{id}(t)) \qquad (6)$$

$$x_{id}(t+1) = x_{id}(t) + v_{id}(t+1). \qquad (7)$$

(The equations are presented for the $d^{th}$ dimension of the position and velocity of the $i^{th}$ particle.)

Here, $c_1$ and $c_2$ are two positive constants, called cognitive learning rate and social learning rate respectively, $\varphi_1$ and $\varphi_2$ are two random functions in the range [0,1], $\omega$ is the time-decreasing inertia factor designed by Eberhart and Shi [9].

The inertia factor balances the global wide-range exploitation and the nearby exploration abilities of the swarm.

### 4. APPLICATION OF THE PSO ALGORITHM TO ACHIEVE AN INTEGRAL APPROXIMATION OF THE FRACTIONAL SYSTEM

We have considered a fractional order process whose transfer function is of the form $\dfrac{1}{a_1 s^\alpha + a_2 s^\beta + a_3}$. We want to have an integer order approximate model of this fractional order process. For the present problem, we assume that $a_1$, $a_2$, $a_3$, $\alpha$ and $\beta$ are known, $\alpha$ and $\beta$ being fractional.

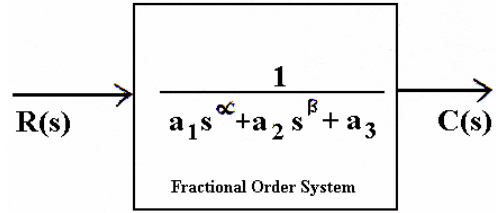

Fig. 1. System for which We Need an Integral Order Model

We apply $R(s)=1/s$ (unit step) and $R(s)=1/s^2$ (unit ramp) to the fractional order system and obtain sampled values of output $c_1(t)$ and $c_2(t)$ respectively. The PSO algorithm will search the solution space to come up with an integral order process model which replicates the observed $c_1(t)$ and $c_2(t)$ values for the same input signals.

Let $n_1 < \alpha < n_1 + 1$ and $n_2 < \beta < n_2 + 1$ where $n_1$ and $n_2$ are integers. We can then assume that $\dfrac{1}{a_1 s^\alpha + a_2 s^\beta + a_3}$ may be approximated by $\dfrac{1}{b_1 s^{n_1+1} + b_2 s^{n_1} + b_3 s^{n_2+1} + b_4 s^{n_2} + b_5}$.

For the present case $a_1=0.8$, $a_2=0.5$, $a_3=1$, $\alpha=2.2$ and $\beta=0.9$. Hence $n_1=2$ and $n_2=0$. Therefore, $\dfrac{1}{0.8s^{2.2} + 0.5s^{0.9} + 1}$ can be approximated as $\dfrac{1}{b_1 s^3 + b_2 s^2 + b_3 s + b_4}$ where $b_1$, $b_2$, $b_3$ and $b_4$ are the unknown coefficients whose values we have to determine.

Since there are four unknowns, namely $b_1$, $b_2$, $b_3$ and $b_4$; the solution space is four-dimensional. So each particle of the population has a four-dimensional position vector ($b_1$, $b_2$, $b_3$, $b_4$). The velocity vectors, the personal bests and the global best are also four-dimensional. The limits on the position vectors of the particles are set by us as follows. As a practical case, we assume that the $b_1$, $b_2$, $b_3$ and $b_4$ all lie between 0 to 2. i.e. $0 \leq b_1, b_2, b_3, b_4 \leq 2$.

The inertia factor $\omega$ decrease linearly from 0.9 to 0.4. The cognitive learning rate $c_1=1.4$. The social learning rate $c_2=1.4$. Number of particles in the population is twenty.

Initially there are twenty solution sets of $\{b_1, b_2, b_3, b_4\}$ where $b_1, b_2, b_3, b_4$ lie in the range 0 to 2 randomly. Thus we have twenty random process models. We consider application of unit step and unit ramp input to each of these process models and obtain by numerical inverse laplace, two sets of outputs, one for unit step and another for unit ramp input. Let us call these $p_1(t)$ and $p_2(t)$ respectively.

We find $F_1 = \sum_t [c_1(t) - p_1(t)]^2$ (for unit step input) and $F_2 = \sum_t [c_2(t) - p_2(t)]^2$ (for unit ramp input). We the find F = $F_1 + F_2$, which is the sum of square deviations from the set of observations for the actual fractional order process. F is the fitness function which the PSO algorithm attempts to minimize, the minimum value evidently being zero. At F = 0, the unknown parameters are optimized.

After running the PSO algorithm, we obtain the position vector of the best particle, i.e. the optimized values of $\{b_1, b_2, b_3, b_4\}$.

This optimized solution set gives the estimated parameter set. The integral order process model corresponding to this optimized solution set should provide outputs almost identical to $c_1(t)$ and $c_2(t)$ for unit step and unit ramp inputs respectively.

## 5. ILLUSTRATION

The process that is to be approximated is $\dfrac{1}{0.8s^{2.2} + 0.5s^{0.9} + 1}$. Synthetic data for $c_1(t)$ [input: r(t) = 1] and $c_2(t)$ [input: r(t) = t] are created, i.e. the values of $c_1(t)$ and $c_2(t)$ are obtained at different time instants assuming a process with transfer function $\dfrac{1}{0.8s^{2.2} + 0.5s^{0.9} + 1}$.

Sampling frequency is 20 samples per second.

Now the PSO algorithm searches for the optimum set of parameters $\{b_1, b_2, b_3, b_4\}$ for which the approximate process model $\dfrac{1}{b_1 s^3 + b_2 s^2 + b_3 s + b_4}$ has unit step and unit ramp responses with minimum deviation from $c_1(t)$ and $c_2(t)$. The best estimations of the process parameters are $b_1 = 0.1772, b_2 = 0.7329, b_3 = 0.4463, b_4 = 1.0265$.

So the integer order approximation is $\dfrac{1}{0.1772s^3 + 0.7329s^2 + 0.4463s + 1.0265}$.

The best integer order process model has square error 0.3788 (considering deviations from $c_1(t)$ and $c_2(t)$ both).

The unit step and unit ramp responses of the actual and the approximate systems are given below.

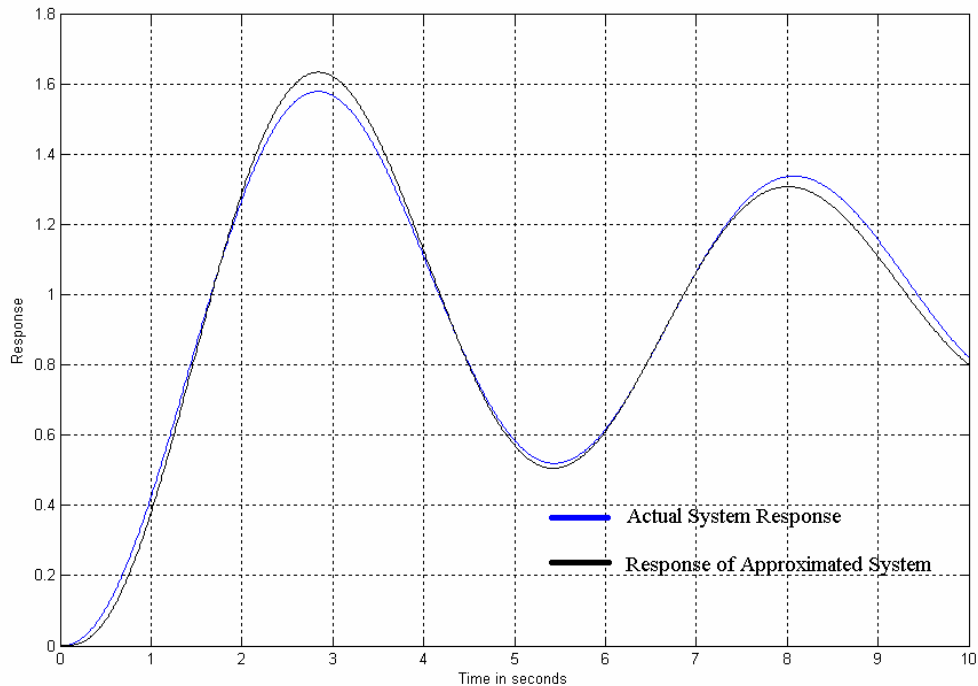

Fig. 2. Unit Step Responses of Actual and Approximate Systems

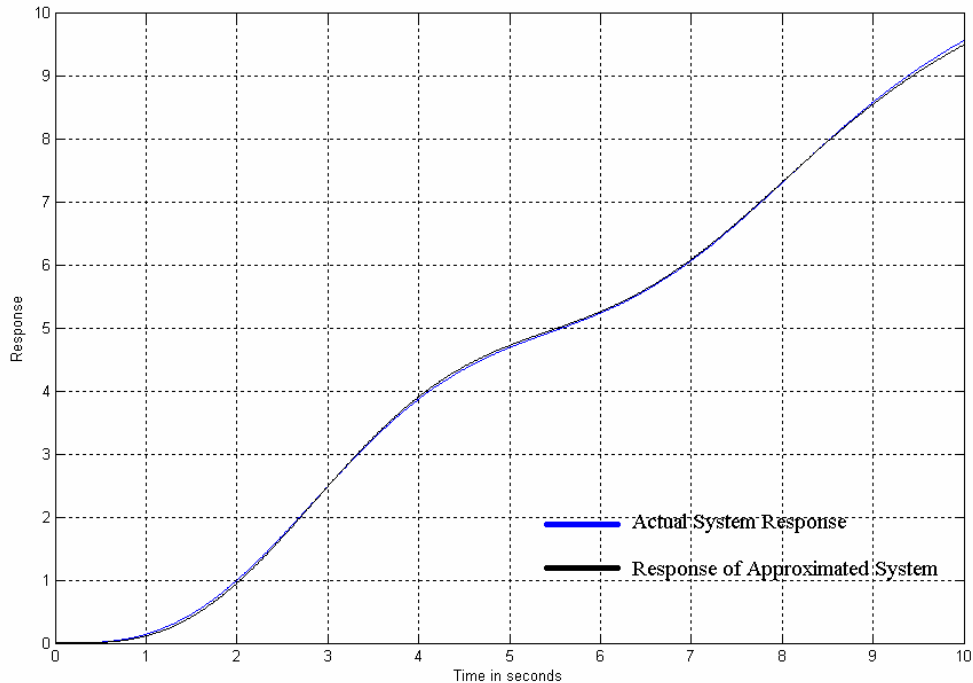

Fig. 3. Unit Ramp Responses of Actual and Approximate Systems

## 6. CONCLUSIONS

An elegant method for the integral order approximation of a fractional order system is proposed. The proposed method provides quite accurate results in spite of a low sampling rate. For usual industrial processes, the slight inaccuracy can be tolerated. The process of approximation can actually be implemented by a simple computer program. The same method can also be used to obtain a simple approximate model of a complex process.